\documentstyle[aps,preprint]{revtex}
\begin{document}
\draft
\tightenlines
 
\title{Reweighting in Nonequilibrium Simulations}
\author{Ronald Dickman$^{1,2}$}
\address{
Departamento de F\'{\i}sica,
Universidade Federal de Santa Catarina,
Campus Universit\'ario\\
Trindade, CEP 88040-900,
Florian\'opolis --- SC, Brasil\\}
\date{\today}

\maketitle
\begin{abstract}
A simple reweighting scheme is proposed for Monte Carlo simulations
of interacting particle systems, permitting one to study 
various parameter values in a single study, and
improving efficiency by an order of magnitude.
Unlike earlier reweighting schemes, the present approach does not require
knowledge of the stationary probability distribution, and so is
applicable out of equilibrium.
The method is applied to the contact process in two and three
dimensions, yielding the critical parameter and spreading exponents to
unprecedented precision.
\end{abstract}


\newpage

Reweighting and histogram methods, which permit one to calculate
thermal averages at different temperatures from a single sample,
have greatly improved the efficiency of Monte Carlo
simulations \cite{ferr}.
Given a sample of configurations, $C_1,...,C_N$, generated at
temperature $T$, one can in principle generate a sample
appropriate to some temperature $T'$, by assigning 
the weight $w_i$ = $P_i (T')/P_i (T)$ to configuration $i$.
Until now, such methods have been
restricted to equilibrium, where the
stationary probability distribution is $P_i \propto e^{-E_i/kT}$.
Away from equilibrium, one does not in general know the stationary
distribution, so there is no simple way to evaluate
the $w_i$.  Recently Grassberger and Zhang showed how a
``self-organized" formulation of directed percolation can be used 
to study several parameter values in a 
single run, without reweighting \cite{gz}.
In this work I introduce a reweighting scheme for
interacting particle systems, based on the
observation that, despite our ignorance of the {\it stationary}
distribution on configuration space, we can write down the probability for any
sequence of events between time zero and time $t$.  
 
I apply the reweighting method to the contact process (CP),
a simple particle system (lattice Markov process)
exhibiting a phase transition to an absorbing
(frozen) state at a critical value of the creation rate \cite{liggett}.  This model belongs to the
universality class of directed percolation \cite{kinzel} and Reggeon field theory \cite{rft},
(it is one of the most well-studied
representatives of this class \cite{rev}),
and is pertinent to models of epidemics \cite{harris}, 
catalysis \cite{catal}, 
and damage spreading \cite{damage}.
In the CP
each site of the hypercubic lattice ${\cal Z}^d$ 
is either vacant or occupied by a particle.
Particles are created at vacant sites at rate $\lambda n /2d$, where $n$ is
the number of occupied nearest-neighbors, and
are annihilated at unit rate, independent of the surrounding
configuration. The order parameter is the particle
density $\rho$; the vacuum state, $\rho = 0 $ is absorbing.
As $\lambda$ is increased beyond $\lambda_c$,
there is a continuous phase transition from the vacuum 
to an active state; for $\lambda > \lambda_c$, 
$\rho \sim (\lambda - \lambda_c)^{\beta}$ in the stationary state.

There are a number of ways (equivalent as regards scaling behavior), 
of implementing the CP in a simulation algorithm;
in this work I follow the widely used practice of maintaining a list
of all occupied sites.  Trials begin at time zero, from a fixed initial
configuration.  Subsequent events involve selecting (at random) an occupied
site {\bf x} from the $N_p$ sites on the list, selecting a process --- creation with probability
$p = \lambda/(1+\lambda)$, annihilation with probability $1-p$ --- and, in the
case of creation, selecting one of the $2d$ nearest neighbors, {\bf y}, 
of {\bf x}.  (The creation attempt succeeds if {\bf y} is vacant).
The time increment $\Delta t$ associated with an event is $1/N_p$, where $N_p$
is the number of occupied sites immediately prior to the event.
A trial ends when all the particles have vanished, or at the
first event with time $\geq t_m$, a predetermined maximum time. 

Consider a single trial, extending from time zero
up to $t_m$; for simplicity, suppose that initially there is but a
single particle, located at the origin.  (If all of the particles disappear
at some time $t'$, the trial is trapped in the vacuum state
for all later times.)
A trial consists of sequence $S$ of events,
each involving the annilihation or (attempted) creation of a particle.
With the help of diagrams reminiscent of the ``percolation substructure"
invoked in defining the CP \cite{griffeath,durrett}, we can write down the
probability of sequence $S$; examples are shown in Fig. 1.  Each annihilation
event carries a factor of $(1-p)/N_p$, each creation event (succesful or not)
a factor $p/(2dN_p)$.
The probability $P_p(S)$ of a sequence $S$, extending to time $t$,
is simply the product
of all factors associated with events occuring at times $t' \leq t$.
Now, for finite $t$, the set ${\cal S}_t$ of possible sequences is finite,
and if $A(S;t)$ is
any property of the system (e.g., the number of particles at time $t$),
then its expectation is

\begin{equation}
\langle A_t \rangle_p =  \sum_{S \in {\cal S}_t} P_p(S) A(S;t) \;.
\label{expa}
\end{equation}

In a Monte Carlo simulation we generate a sample $S_1$,...,$S_N$, drawn
from the distribution $P_p(S)$, which yields the estimate

\begin{equation}
\overline{A}_{t;p} \equiv  \frac{1}{N} \sum_{k=1}^N A(S_k;t) \;.
\label{esta}                          
\end{equation}
From our analysis of $P_p(S)$, it is evident that its $p$-dependence
only involves the numbers $c(S)$ and $a(S)$ of creation and annihilation
events, respectively; the ratio of the probabilities
associated with two different values of $p$ is

\begin{equation}
\omega (S) \equiv \frac{P_{p'}(S)}{P_p(S)} =
\left( \frac{p'}{p} \right)^{c(S)}   \left( \frac{1-p'}{1-p} \right)^{a(S)} \;.
\label{omega}
\end{equation}
Thus the reweighted estimate,

\begin{equation}
\overline{A}_{t;p'} \equiv  \frac{1}{N} \sum_{k=1}^N \omega (S_k) A(S_k;t) \;,
\label{estap}
\end{equation}
has expectation
\begin{equation}
\sum_{S \in {\cal S}_t} P_p(S) \omega (S) A(S;t)
                               = \langle A_t \rangle_{p'},
\end{equation}
and is an unbiased estimator of $A$ in the process with
creation probability $p'$.

As in other applications of reweighting, it is not enough to have an
unbiased estimator; one must also ensure that the sample generated with
creation probability $p$ has a reasonable degree of overlap with a
typical sample at $p'$.  We expect that as the duration $t_m$ increases,
the range of $\lambda$ values for which a sample is useful will narrow.  To
estimate this range of values, consider a sequence of
$r$ events.  The probability that exactly $c$ of these are creation
attempts is given by the binomial distribution,

\begin{equation}
P(c;r) = \frac {r!}{c! (r-c)!} \frac {\lambda^c}{(1+\lambda)^r}
\end{equation}
so that $\langle c \rangle = rp $ and the standard deviation
$\sigma$ is given by $\sigma^2 = r \lambda/(1+\lambda)^2$.  In a typical
sequence generated with creation probability p, the actual number of
creation events will be in the range $rp \pm \sigma$, corresponding to
a creation rate of $\lambda' \simeq \lambda[1 \pm (1+\lambda)(\lambda r)^{-1/2}]$.
Thus the typical fluctuation in $\lambda$ is

\begin{equation}
\delta \lambda  = (1+\lambda) \sqrt{\frac{\lambda}{r}}.
\label{dellam}
\end{equation}
In the two-dimensional CP, for example, one finds that $r \sim 0.7 \; t^{1.68}$
for $\lambda \approx \lambda_c \simeq 1.6488$.  This corresponds to
$\delta \lambda = 0.013$ for $t_m=1000$.   Although the range may
appear narrow, it is more than sufficient for 
``time-dependent" simulations, which typically focus on a
small interval near $\lambda_c$. 

I have applied reweighting to the CP in two and three dimensions,
in studies using two kinds of initial states.  In the case described 
above, one
studies the survival and spread of activity from a single
`seed' particle.  
(The system size must be sufficient to guarantee that particles never
reach the boundaries.)  
The principal quantities of interest are
the survival probability $P(t)$, the mean number of particles $n(t)$,
and the mean-square distance $R^2$ of particles from the origin.
At the critical point these are known to follow asymptotic
power-laws \cite{torre},

\begin{equation}
P(t) \sim t^{-\delta}  \;,
\label{delta}
\end{equation}

\begin{equation}
n(t) \sim t^{\eta}   \;,
\label{eta}
\end{equation}
and
\begin{equation}
R^2(t) \sim t^{z}   \;.
\label{z}
\end{equation}

\noindent Away from $\lambda_c$ these quantities show deviations from power laws.
Since the CP exhibits corrections to scaling of
the form $P(t) \sim t^{-\delta} [ 1 + a t^{-\alpha} + \cdots]$
(similarly for $n$ and $R^2$), it is useful to plot derivatives
(local slopes of the log-log plots) versus $t^{-1}$ when extracting the critical
exponents $\delta$, $\eta$ and $z$ \cite{torre}.
These exponents are connected by the hyperscaling relation:
$4\delta + 2 \eta = d z$.

The second kind of study begins with all sites occupied, and follows
the decay of the particle density $\rho (t)$.  In this case the
signature of the critical point is power-law decay,
$\rho \sim t^{-\theta}$ in the short-time regime, i.e., before the
correlation length has attained the system size $L$.  A scaling
argument implies $\theta = \delta$ \cite{abw,mgt}.

As a preliminary test, I compared two results for
the survival prbability $P(t)$ in the two-dimensional
CP at $\lambda = 1.665$, one obtained by reweighting a sample of $N =10^5$
trials generated at $\lambda = 1.650$, the other
from a similar run but without reweighting ($\lambda = 1.665$). 
The relative difference between the two results for $P(t)$ remains
$\leq 0.017$ for $t \leq t_m = 400$.  Since
each result has a relative uncertainty of 
$\sqrt{(1-P(t))/NP(t)} \simeq 0.01$ for $t = t_m$, ($P(t_m) \simeq 0.085$),
the difference between the two results 
is fully consistent with sample-to-sample fluctuations.

In two dimensions I performed spreading simulations extending to $t_m =2980$,
on lattices of up to $1200 \times 1200$ sites.  Samples 
generated at a central value, $\lambda_0$,
were reweighted so as to study ten additional values, 
$\lambda = \lambda_0 \pm m \Delta \lambda$, with $m = 1,...,5$.  The general 
strategy is to use relatively small samples and run times initially, to bracket $\lambda_c$,
and then extend the sample size and run time to make finer distinctions.
Thus a sample of 10$^6$ trials with $t_m = 665$ and $\Delta \lambda = 10^{-4}$
is already sufficient to restrict $\lambda_c $ to the interval
[1.64875, 1.64895].
The most sensitive indicator of
criticality is the local-slope plot of $\eta_t$,
defined as the derivative of $\ln n$ with
respect to $\ln t$.
[Numerically $\eta_t$ is estimated from a least-squares linear fit
to the $\ln n$ data for a set of $n_i = $ 17-25 equally-spaced values (an increment of
0.1) of $\ln t$;
it is plotted versus $t_a^{-1}$, $t_a$ being the geometric mean of the 
$t$-values over the $n_i$ intervals.  $\delta_t$ and $z_t$ are obtained
similarly.]

To refine the estimate for $\lambda_c$ I generated two samples of
$2.5 \times 10^7$ trials each, with $t_m =2980$, $\lambda_0 = 1.64880$, and
$\Delta \lambda = 4 \times 10^{-5}$.
The critical point was determined from the plot of $\eta_t $,
which shows a clear deviation from smooth behavior for off-critical
values.  (Fig. 2 shows the $\eta_t$ plot for one of the two runs.
Note that the leading correction to scaling does not follow a simple
$1/t$ decay, as was also noted for DP in 2+1 dimensions \cite{gz}.)
These studies yield $\lambda_c = 1.64877(3)$, the number in parentheses
denoting the uncertainty in the last figure.
Extrapolating the local-slope plots, I obtain
$\delta = 0.4523(10)$, $\eta = 0.2293(4)$, and $z = 1.1316(4)$,
which are in good agreement with, and generally sharper than,
the results of previous large-scale simulations of directed
percolation \cite{gz}, and of the ZGB surface catalysis model \cite{vz}
(see Table I).

Grassberger and Zhang \cite{gz} noted that the derivatives of $\ln n(t)$ and $\ln P(t)$
with respect to $\lambda$ (evaluated at $\lambda_c$), grow
$\sim t^{1/\nu_{||}}$.  Analyzing $n(t)$ in this fashion yields
$\nu_{||} = 1.292(4)$, in good agreement with their estimate of 1.295(6).

In three dimensions, I performed four runs
of $10^7$ trials each, extending to $t_m = 2208$.
(To avoid finite-size effects, I do not use an occupancy array
in the three dimensional simulations,
but simply search the particle list to detect overlaps.)
Two studies used $\lambda_0 = 1.31686$,
$\Delta \lambda = 3 \times 10^{-5}$; in the others,
$\lambda_0 = 1.31689$ and $\Delta \lambda = 2 \times 10^{-5}$.
Fig. 3 shows the local
slopes for one of the latter runs.  While the extrapolation of
$\eta$ is straightforward, the strong linear correction to $\delta$
($\delta_t \simeq \delta - 4.96 t^{-1} + \cdots  $), renders it advantageous
to subtract the linear term when estimating the exponent (inset of Fig. 3b).
In the case
of $z$, the curves for all $\lambda$ values are virtually identical.  It is
difficult to extrapolate the $1/t$ plot; the present estimate is
derived by extrapolating the local slope plotted versus $t^{-1/4}$,
as in Fig. 3c.  
Averaging over the results for the four runs yields the following
estimates for the three dimensional CP:
\[
\lambda_c = 1.31686(1),  \;\;\;\;\;\;\;\;
\eta = 0.110(1), 
\]
\[
\delta = 0.7263(11),      \;\;\;\;\;\;\;\;\;
z = 1.042(2) 
\]
(the uncertainties represent one standard deviation).  
These are in good accord with
hyperscaling: $4\delta + 2\eta - 3 z = -0.001(12)$.
The present results are compatible with Jensen's estimates of
$\lambda_c = 1.3168(1)$,
$\eta = 0.114(4)$, and $\delta = 0.730(4)$, but not with his
value $z = 1.052(3)$ \cite{iwan3d}.   Analysis of $d \ln n(t)/d\lambda$
yields $\nu_{||} = 1.114(4)$, just consistent with Jensen's result,
1.105(5).
  
I studied the initial decay of the density (starting with all sites occupied),
in the two-dimensional CP for system sizes $L = 32$, 64, and 128.
Fig. 4 shows the results of one of three sets with $L=128$, $10^5$ trials,
$\lambda_0 = 1.6490$, and $\Delta \lambda = 10^{-4}$.
(The studies for $L=32$ and $64$ used similar parameters, in runs of
$5 \times 10^5$ and
$2 \times 10^5$ trials, respectively.)  
For the relatively short runs employed here ($t_m \leq 1096$), all trials 
survive up to $t_m$.  While the
$\rho(t)$ curves (inset) for different $\lambda $ values are
indistinguishable, the local slopes (main graph) vary quite systematically
with $\lambda$.
Since the power law $\rho \sim t^{-\delta} $ obtains for finite times not
$t \rightarrow \infty$, I plot the
local slope versus $\ln t$ in this case.  One can distinguish three 
regimes: an initial phase in which $\delta_t$ increases for all $\lambda$
values, a late stage in which it decreases, as $\rho (t)$ approaches a
value $\sim L^{-\beta/\nu_{\perp}}$ as predicted by finite size scaling, and
an intermediate regime in which $\delta_t$ is more or less constant.
Associating $\lambda_c$ with the $\delta_t$ most nearly constant in the
intermediate regime yields $\lambda_c$ = 1.6492(1) for $L=32$, 1.6491(1)
for $L=64$, and 1.64898(5) for $L=128$.  The corresponding estimates for
$\delta$ (from the flat portion of each curve) are 0.4508(10) for $L=32$
and 0.4520(5) for $L= 64$ and 128.  Thus the $\lambda_c$ estimates appear to
be approaching the value derived from spreading simulations; the two kinds of
studies yield consistent results for $\delta$.

In summary, I propose a simple reweighting scheme for nonequilibrium
lattice models, and apply it to the contact
process in two and three dimensions.
Since spreading simulations are usually repeated for five or so different
$\lambda$ values, reweighting yields roughly an order-of-magnitude speedup.
In addition to improving efficiency, using the same sample 
to study all parameter values eliminates the effects of independent 
fluctuations, which complicate determination of $\lambda_c$ and the
critical exponents.  
These computational advantages have permitted determination of
the critical parameters of the three-dimensional CP to unprecedented
precision.
One may expect reweighting to find wide
application in simulations of nonequilibrium critical phenomena.
\vspace{1em}

\noindent Acknowledgements
\vspace{1em}

\noindent I thank Adriana Gomes Dickman for suggesting the
initial density decay study, and Miguel Angel Mu\~noz and Robert Ziff
for helpful comments.
\vspace{1em}
 
\noindent $^1${\small electronic address: dickman@fisica.ufsc.br } \\
$^2${\small On leave of absence from: Department of Physics and Astronomy,
Herbert H. Lehman College, City University of New York,
Bronx, NY, 10468-1589.} \\

\begin{table}
\caption{\sf Critical exponents for DP in two dimensions.}
\begin{center}
\begin{tabular}{|l|l|l|l|} 
exponent & Ref. \cite{gz} &  Ref. \cite{vz} & Pres. work  \\
\hline\hline
$\delta$ & 0.451(3) & 0.4505(10) & 0.4523(10) \\
$\eta$   & 0.229(3) & 0.2295(10) & 0.2293(4)  \\
$z$      & 1.133(2) & 1.1325(10) & 1.1316(4)  \\
$4\delta + 2\eta - 2z$ & -0.004(22) & -0.004(8) & 0.005(6)
\end{tabular}
\end{center}
\label{exp2}
\end{table}


\noindent FIGURE CAPTIONS
\vspace{1em}

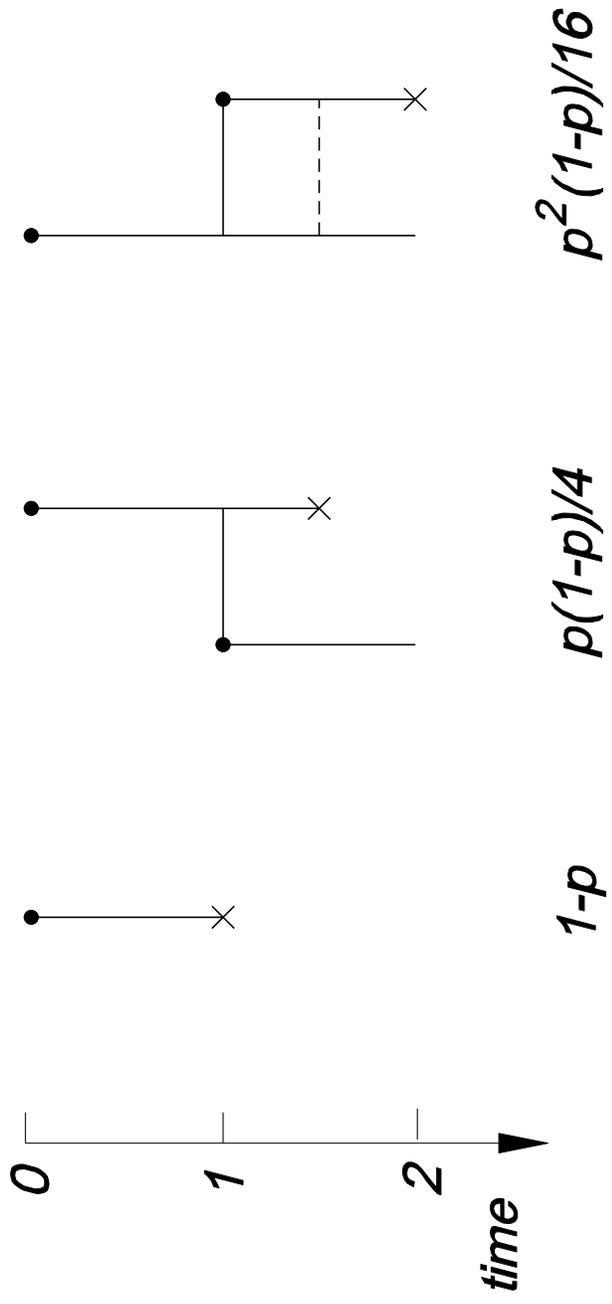
\noindent FIG. 1. Examples of event sequences in the one-dimensional contact
process starting from a single particle.  Vertical lines represent particles,
whose birth is marked by a dot, and annihilation by $\times$.  Solid and
dashed horizontal lines represent, respectively, successful
and unsuccessful creation events.  Probabilites are listed beneath each sequence.
\vspace{1em}

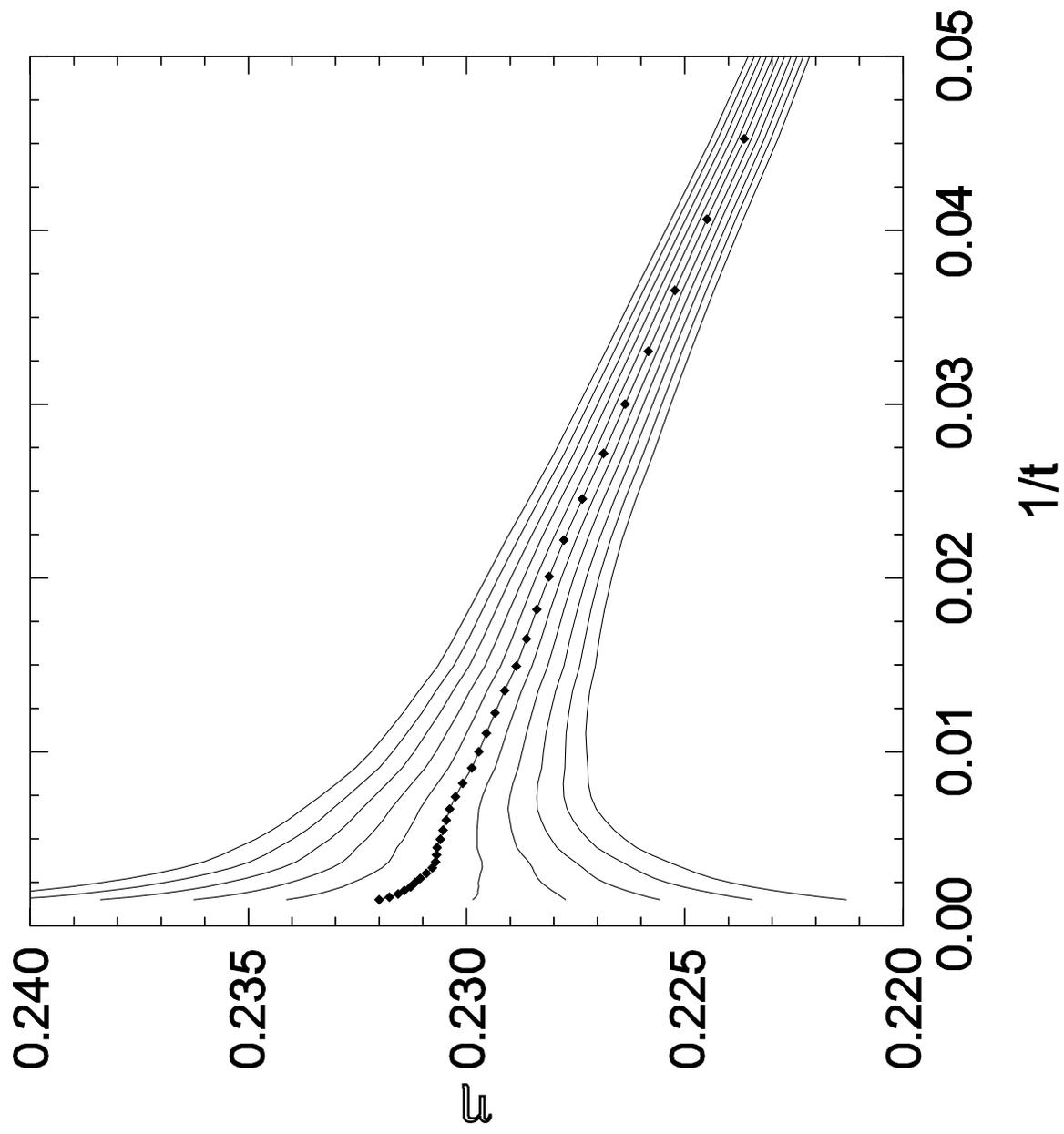
\noindent FIG. 2. Local slope plot for exponent $\eta$ in the two dimensional
CP.  The middle curve (with data points) marks $\lambda_0 = 1.64880$.
Curves above and below are for $\lambda$ values at intervals of
$\Delta \lambda = 4 \times 10^{-5} $ above and below $\lambda_0$.
\vspace{1em}

\noindent FIG. 3a. Local slope plot for exponent $\delta$ in the three
dimensional CP.  Inset: detail of $\delta - 4.96 t^{-1} $.
The middle curve (with data points) marks $\lambda_0 = 1.31689$;
curves above and below are for $\lambda$ values at intervals of
$\Delta \lambda = 2 \times 10^{-5} $.
\vspace{1em}

\noindent FIG. 3b. Local slope plot for exponent $\eta$ in the three
dimensional CP.  Symbols as in Fig. 3a.
\vspace{1em}

\noindent FIG. 3c. Local slope plot for exponent $z$ in the three
dimensional CP.  The inset shows the same data plotted versus $t^{-1/4}$.
\vspace{1em}

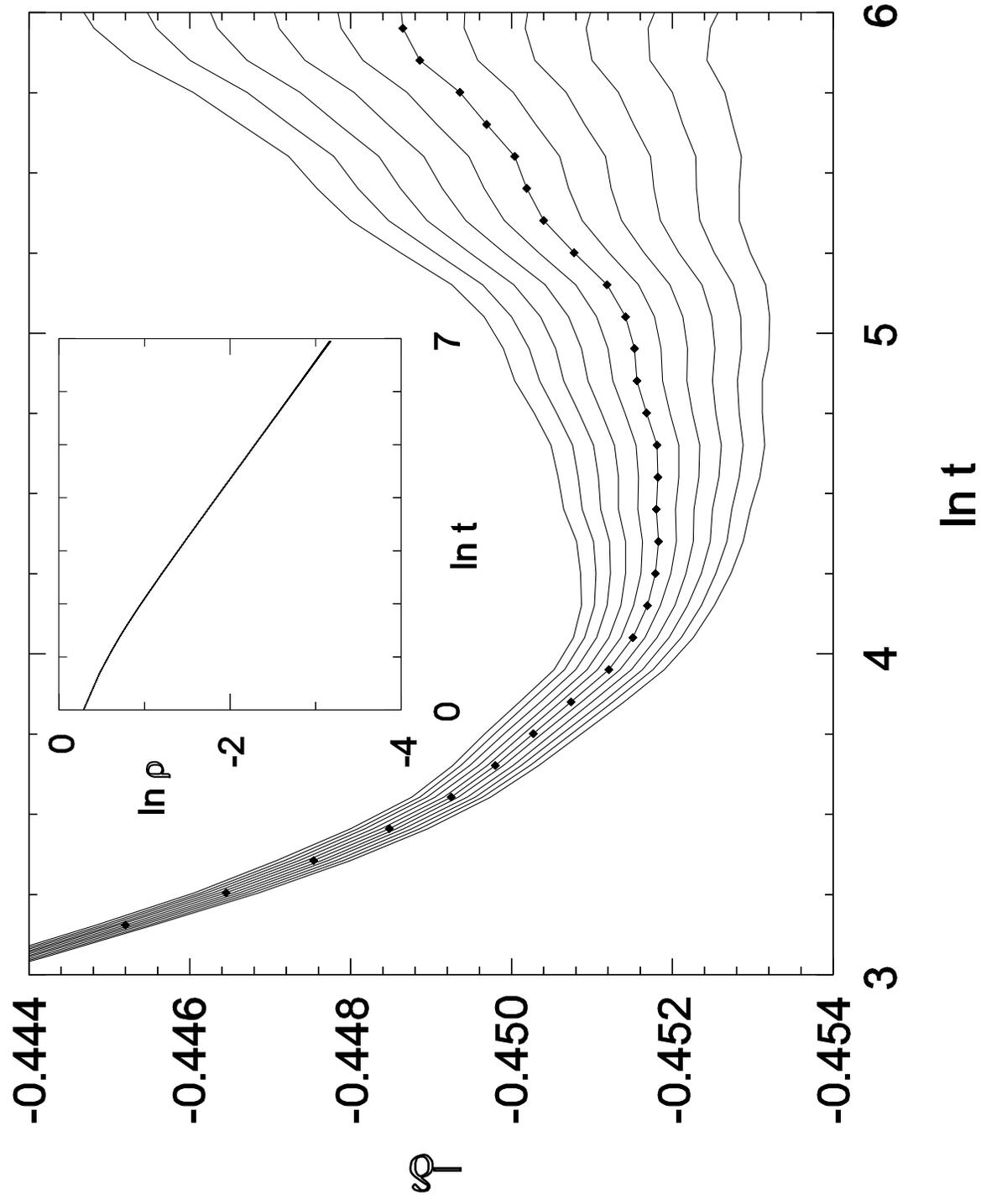
\noindent FIG. 4. Local slope plot for exponent $\delta$ in the two
dimensional CP starting with all sites occupied, $L=128$.   The data
points mark $\lambda_0 = 1.6490$, $\Delta \lambda = 5 \times 10^{-5}$.
Inset: decay of the density $\rho$.

\begin{thebibliography}{99}

\bibitem{ferr}
         A. M. Ferrenberg and R. H. Swendsen,
         Phys. Rev. Lett. {\bf 61}, 2635 (1988);
         Phys. Rev. Lett. {\bf 63}, 1195 (1989).

\bibitem{gz}
         P. Grassberger and Y. Zhang, 
         Physica A{\bf 224}, 169 (1996). 

\bibitem{liggett}
         T. M. Liggett,
         {\it Interacting Particle Systems},
         Springer-Verlag, (New York, 1985).

\bibitem{kinzel}
       W. Kinzel,
       Z. Phys. B{\bf 58}, 229 (1985).
 
\bibitem{rft}
        P. Grassberger,
        Z. Phys. B {\bf 47}, 365 (1982);
        H.K. Janssen,
        Z. Phys. B {\bf 42}, 151 (1981).

\bibitem{rev}
        J. Marro and R. Dickman,
        {\it Nonequilibrium Phase Transitions in Lattice Models},
        Cambridge University Press, (Cambridge, 1999).
        R. Dickman,
        in {\it Nonequilibrium Statistical Mechanics in One Dimension},
        Ed. V. Privman,
        Cambridge University Press, (Cambridge, 1997). 

\bibitem{harris} T.E. Harris, Ann. Phys. {\bf 2}, 969 (1974).
                                                                            
\bibitem{catal} 
        G. Grinstein, D.-W. Lai, and D. A. Browne,
        Phys. Rev. A{\bf 40}, 4820 (1989). 
 
\bibitem{damage}
        P. Grassberger,
        J. Stat. Phys. {\bf 79}, 13 (1995).

\bibitem{griffeath}
         D. Griffeath,  {\it Additive and Cancellative Interacting Particle Systems},
         Lecture Notes in Mathematics v. 724,
         Springer-Verlag, (Berlin, 1979).

\bibitem{durrett}
         R. Durrett,  {\it Lecture Notes on Particle Systems and
         Percolation}, Wadsworth \& Brooks/Cole, (Pacific Grove, CA, 1988).
  
\bibitem{torre}
         P. Grassberger and A. de la Torre,
         Ann. Phys. (N.Y.) {\bf 122}, 373 (1979).

\bibitem{abw}
      T. Aukrust, D. A. Browne, and I. Webman, 
      Phys. Rev. A{\bf 41}, 5294 (1990). 

\bibitem{mgt}
      M. A. Mu\~noz, G. Grinstein, and Y. Tu,
      Phys. Rev. E{\bf 56}, 5101 (1997).

\bibitem{vz}
      C. A. Voigt and R. M. Ziff,
      Phys. Rev. E{\bf 56}, R6241 (1997).

\bibitem{iwan3d}
      I. Jensen, 
      Phys. Rev. A{\bf 45}, R563 (1992);
      Phys. Rev. A{\bf 46}, 7393 (1992).



\end{thebibliography}
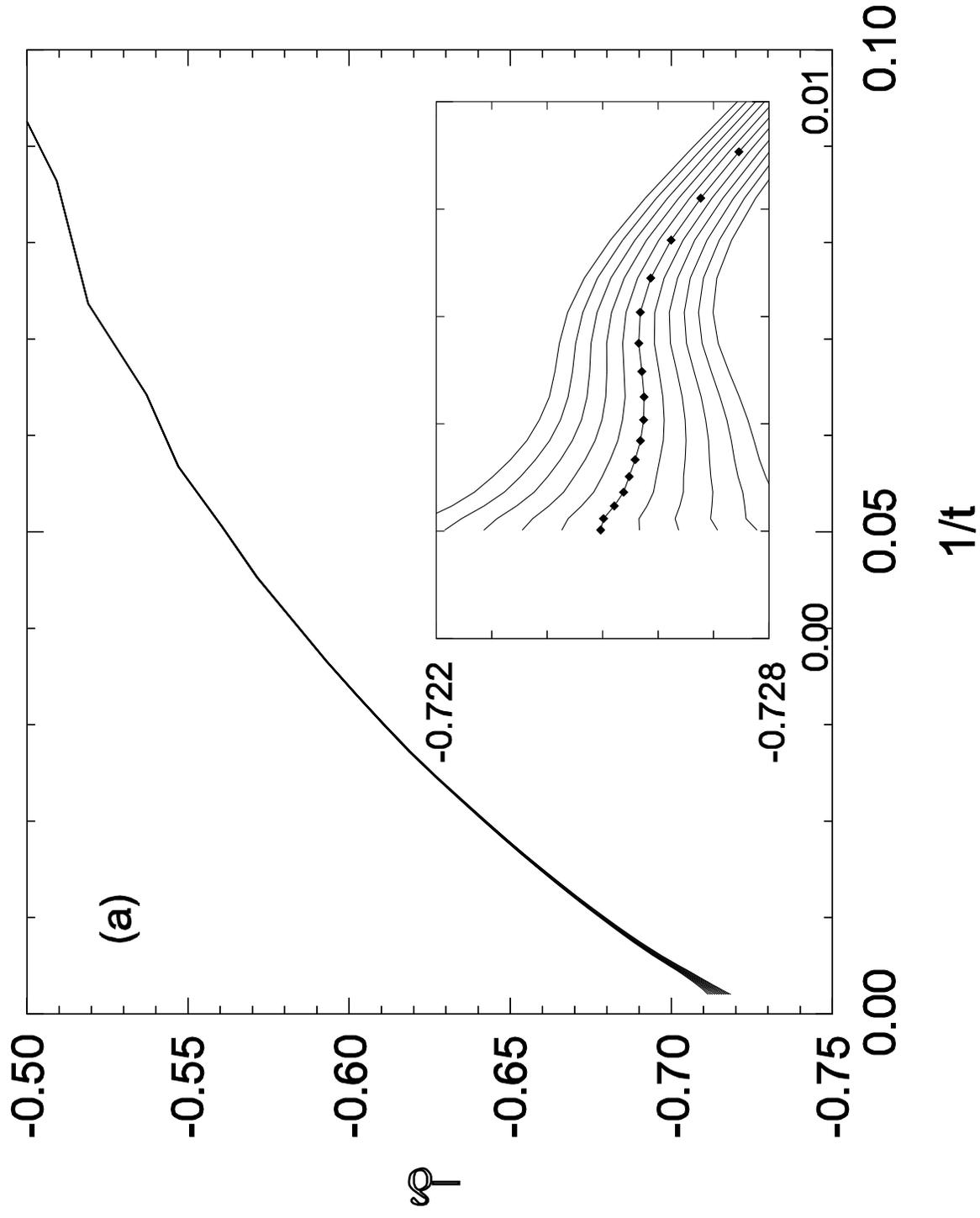
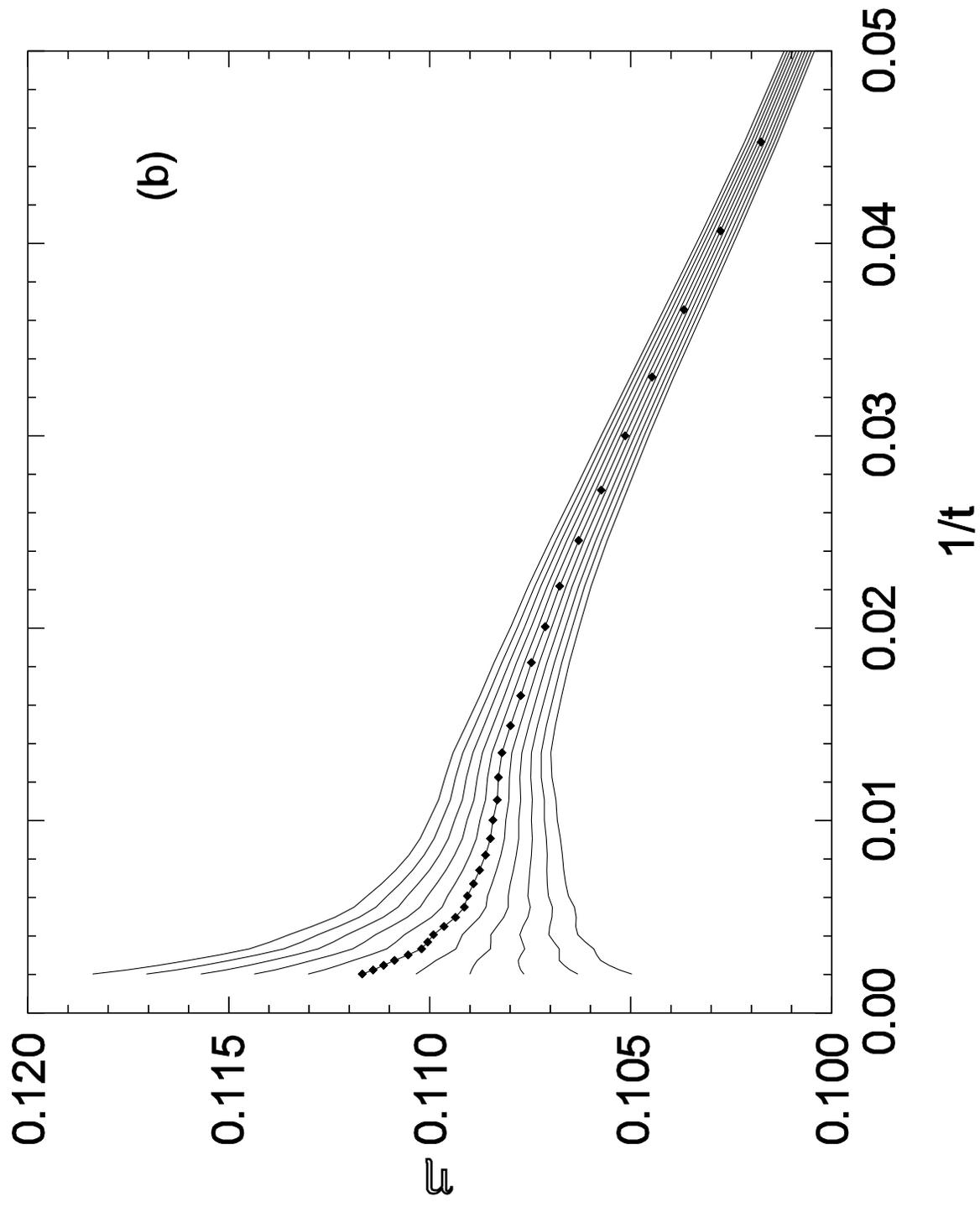
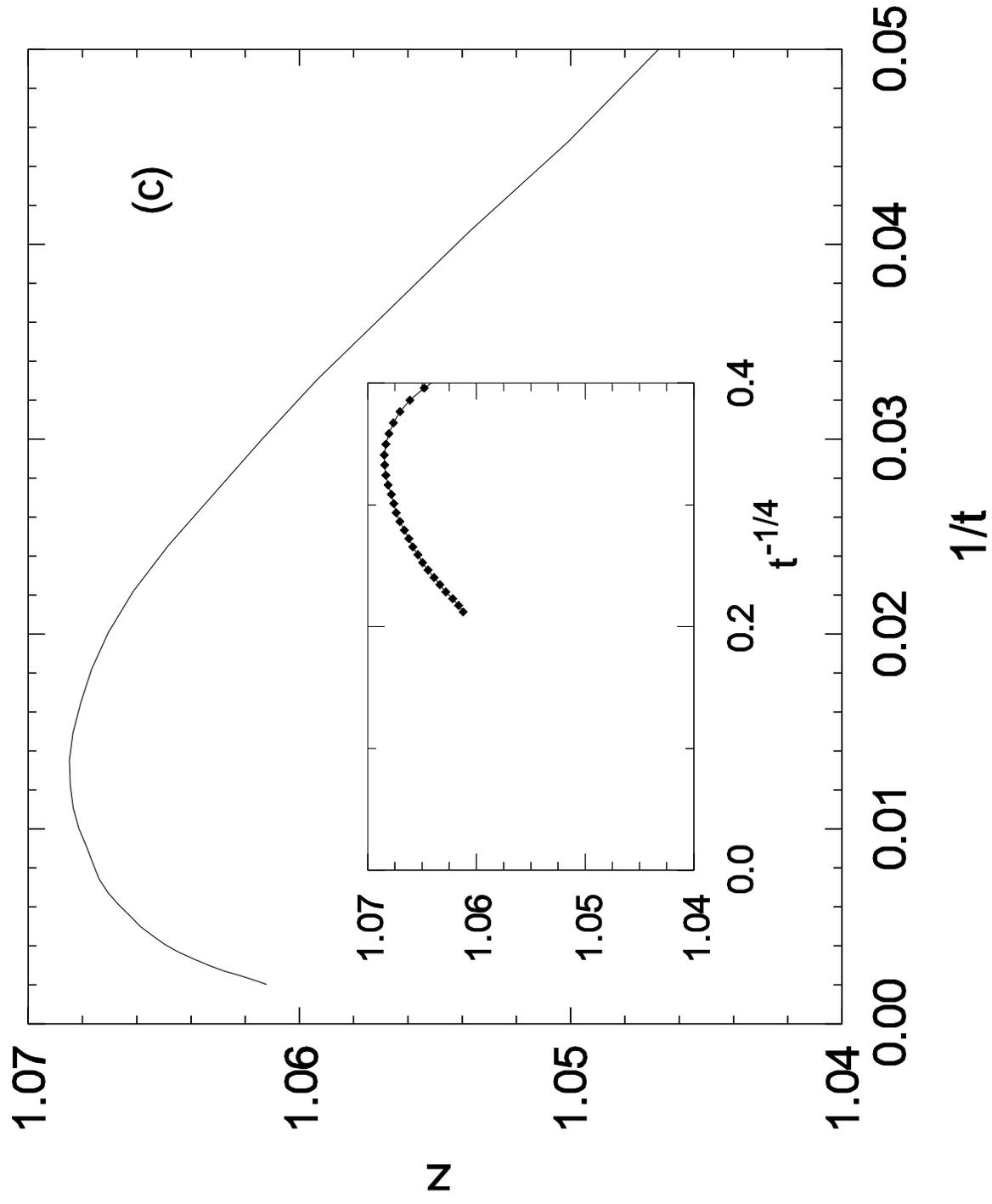
\end{document}